
%
%
%
%
%
%
%
%
\def\standardrisposta{s }\def\reducedrisposta{r }
\def\mplarisposta{mpla }\def\zerorisposta{z }
\def\doublerisposta{d }\def\cartarisposta{e }\def\amsrisposta{y }
\newcount\ingrandimento \newcount\sinnota \newcount\dimnota
\newcount\unoduecol \newdimen\collhsize \newdimen\tothsize
\newdimen\fullhsize \newcount\controllorisposta \sinnota=1
\newskip\infralinea  \global\controllorisposta=0
\immediate\write16 { ********  Welcome to PANDA macros (Plain TeX,
AP, 1991) ******** }
\immediate\write16 { You'll have to answer a few questions in
lowercase.}
\message{>  Do you want it in double-page (d), reduced (r)
or standard format (s) ? }\read-1 to\risposta
\message{>  Do you want it in USA A4 (u) or EUROPEAN A4
(e) paper size ? }\read-1 to\srisposta
\message{>  Do you have AMSFonts 2.0 (math) fonts (y/n) ? }
\read-1 to\arisposta
%
%
%
%
%
\ifx\risposta\standardrisposta \ingrandimento=1200
\message {>> This will come out UNREDUCED << }
\dimnota=2 \unoduecol=1 \global\controllorisposta=1 \fi
\ifx\risposta\reducedrisposta \ingrandimento=1095 \dimnota=1
\unoduecol=1  \global\controllorisposta=1
\message {>> This will come out REDUCED << } \fi
\ifx\risposta\doublerisposta \ingrandimento=1000 \dimnota=2
\unoduecol=2   \message {>> You must print this in
LANDSCAPE orientation << } \global\controllorisposta=1 \fi
\ifx\risposta\mplarisposta \ingrandimento=1000 \dimnota=1
\message {>> Mod. Phys. Lett. A format << }
\unoduecol=1 \global\controllorisposta=1 \fi
\ifx\risposta\zerorisposta \ingrandimento=1000 \dimnota=2
\message {>> Zero Magnification format << }
\unoduecol=1 \global\controllorisposta=1 \fi
\ifnum\controllorisposta=0  \ingrandimento=1200
\message {>>> ERROR IN INPUT, I ASSUME STANDARD
UNREDUCED FORMAT <<< }  \dimnota=2 \unoduecol=1 \fi
\magnification=\ingrandimento
%
%
%
%
\newdimen\eucolumnsize \newdimen\eudoublehsize \newdimen\eudoublevsize
\newdimen\uscolumnsize \newdimen\usdoublehsize \newdimen\usdoublevsize
\newdimen\eusinglehsize \newdimen\eusinglevsize \newdimen\ussinglehsize
\newskip\standardbaselineskip \newdimen\ussinglevsize
\newskip\reducedbaselineskip \newskip\doublebaselineskip
\eucolumnsize=12.0truecm    
\eudoublehsize=25.5truecm   
\eudoublevsize=6.5truein    
\uscolumnsize=4.4truein     
\usdoublehsize=9.4truein    
\usdoublevsize=6.8truein    
\eusinglehsize=6.5truein    
\eusinglevsize=24truecm     
\ussinglehsize=6.5truein    
\ussinglevsize=8.9truein    
\standardbaselineskip=16pt plus.2pt  
\reducedbaselineskip=14pt plus.2pt   
\doublebaselineskip=12pt plus.2pt    
%
%
\def\Portoffset{}
\def\Landoffset{}
\ifx\risposta\mplarisposta \def\Portoffset{\hoffset=1.8truecm} \fi
%
%
\def\Landspec{}
\tolerance=10000
\parskip=0pt plus2pt  \leftskip=0pt \rightskip=0pt
%
%
\ifx\risposta\standardrisposta \infralinea=\standardbaselineskip \fi
\ifx\risposta\reducedrisposta  \infralinea=\reducedbaselineskip \fi
\ifx\risposta\doublerisposta   \infralinea=\doublebaselineskip \fi
\ifx\risposta\mplarisposta     \infralinea=13pt \fi
\ifx\risposta\zerorisposta     \infralinea=12pt plus.2pt\fi
\ifnum\controllorisposta=0    \infralinea=\standardbaselineskip \fi
\ifx\risposta\doublerisposta   \Landoffset \else \Portoffset \fi
\ifx\risposta\doublerisposta \ifx\srisposta\cartarisposta
\tothsize=\eudoublehsize \collhsize=\eucolumnsize
\vsize=\eudoublevsize  \else  \tothsize=\usdoublehsize
\collhsize=\uscolumnsize \vsize=\usdoublevsize \fi \else
\ifx\srisposta\cartarisposta \tothsize=\eusinglehsize
\vsize=\eusinglevsize \else  \tothsize=\ussinglehsize
\vsize=\ussinglevsize \fi \collhsize=4.4truein \fi
\ifx\risposta\mplarisposta \tothsize=5.0truein
\vsize=7.8truein \collhsize=4.4truein \fi
%
%
%
%
\newcount\contaeuler \newcount\contacyrill \newcount\contaams
\font\ninerm=cmr9  \font\eightrm=cmr8  \font\sixrm=cmr6
\font\ninei=cmmi9  \font\eighti=cmmi8  \font\sixi=cmmi6
\font\ninesy=cmsy9  \font\eightsy=cmsy8  \font\sixsy=cmsy6
\font\ninebf=cmbx9  \font\eightbf=cmbx8  \font\sixbf=cmbx6
\font\ninett=cmtt9  \font\eighttt=cmtt8  \font\nineit=cmti9
\font\eightit=cmti8 \font\ninesl=cmsl9  \font\eightsl=cmsl8
\skewchar\ninei='177 \skewchar\eighti='177 \skewchar\sixi='177
\skewchar\ninesy='60 \skewchar\eightsy='60 \skewchar\sixsy='60
\hyphenchar\ninett=-1 \hyphenchar\eighttt=-1 \hyphenchar\tentt=-1
%
\font\tencmmib=cmmib10  \newfam\cmmibfam  \skewchar\tencmmib='177
\font\tencmbsy=cmbsy10  \newfam\cmbsyfam  \skewchar\tencmbsy='60
\def\scaps{\cmcsc}                 
\font\tencmcsc=cmcsc10  \newfam\cmcscfam
\ifnum\ingrandimento=1095

\font\capsone=cmcsc10 at 10.95pt 

\else

\font\capsone=cmcsc10 at 12pt 
\fi

\def\ttaarr{\bf}		
\def\ppaarr{\sl}		

%
%
%
\newfam\eufmfam \newfam\msamfam \newfam\msbmfam \newfam\eufbfam
\def\Loadeulerfonts{\global\contaeuler=1 \ifx\arisposta\amsrisposta
\font\teneufm=eufm10              
\font\eighteufm=eufm8 \font\nineeufm=eufm9 \font\sixeufm=eufm6
\font\seveneufm=eufm7  \font\fiveeufm=eufm5
\font\teneufb=eufb10              
\font\eighteufb=eufb8 \font\nineeufb=eufb9 \font\sixeufb=eufb6
\font\seveneufb=eufb7  \font\fiveeufb=eufb5
\font\teneurm=eurm10              
\font\eighteurm=eurm8 \font\nineeurm=eurm9
\font\teneurb=eurb10              
\font\eighteurb=eurb8 \font\nineeurb=eurb9
\font\teneusm=eusm10              
\font\eighteusm=eusm8 \font\nineeusm=eusm9
\font\teneusb=eusb10              
\font\eighteusb=eusb8 \font\nineeusb=eusb9
\else \def\eufm{\tt} \def\eufb{\tt} \def\eurm{\tt} \def\eurb{\tt}
\def\eusm{\tt} \def\eusb{\tt}    \fi}

\def\loadamsmath{\global\contaams=1 \ifx\arisposta\amsrisposta
\font\tenmsam=msam10 \font\ninemsam=msam9 \font\eightmsam=msam8
\font\sevenmsam=msam7 \font\sixmsam=msam6 \font\fivemsam=msam5
\font\tenmsbm=msbm10 \font\ninemsbm=msbm9 \font\eightmsbm=msbm8
\font\sevenmsbm=msbm7 \font\sixmsbm=msbm6 \font\fivemsbm=msbm5
\else \def\msbm{\bf} \fi \def\Bbb{\msbm} \def\symbl{\msam} \tenpoint}
\def\loadcyrill{\global\contacyrill=1 \ifx\arisposta\amsrisposta
\font\tenwncyr=wncyr10 \font\ninewncyr=wncyr9 \font\eightwncyr=wncyr8
\font\tenwncyb=wncyr10 \font\ninewncyb=wncyr9 \font\eightwncyb=wncyr8
\font\tenwncyi=wncyr10 \font\ninewncyi=wncyr9 \font\eightwncyi=wncyr8
\else \def\cyrill{\sl} \def\cyrilb{\sl} \def\cyrili{\sl} \fi\tenpoint}
\ifx\arisposta\amsrisposta
\font\sevenex=cmex7               
\font\eightex=cmex8  \font\nineex=cmex9
\font\ninecmmib=cmmib9   \font\eightcmmib=cmmib8
\font\sevencmmib=cmmib7 \font\sixcmmib=cmmib6
\font\fivecmmib=cmmib5   \skewchar\ninecmmib='177
\skewchar\eightcmmib='177  \skewchar\sevencmmib='177
\skewchar\sixcmmib='177   \skewchar\fivecmmib='177
\font\ninecmbsy=cmbsy9    \font\eightcmbsy=cmbsy8
\font\sevencmbsy=cmbsy7  \font\sixcmbsy=cmbsy6
\font\fivecmbsy=cmbsy5   \skewchar\ninecmbsy='60
\skewchar\eightcmbsy='60  \skewchar\sevencmbsy='60
\skewchar\sixcmbsy='60    \skewchar\fivecmbsy='60
\font\ninecmcsc=cmcsc9    \font\eightcmcsc=cmcsc8     \else
\def\cmmib{\fam\cmmibfam\tencmmib}\textfont\cmmibfam=\tencmmib
\scriptfont\cmmibfam=\tencmmib \scriptscriptfont\cmmibfam=\tencmmib
\def\cmbsy{\fam\cmbsyfam\tencmbsy} \textfont\cmbsyfam=\tencmbsy
\scriptfont\cmbsyfam=\tencmbsy \scriptscriptfont\cmbsyfam=\tencmbsy
\scriptfont\cmcscfam=\tencmcsc \scriptscriptfont\cmcscfam=\tencmcsc
\def\cmcsc{\fam\cmcscfam\tencmcsc} \textfont\cmcscfam=\tencmcsc \fi
\catcode`@=11
\newskip\ttglue
\gdef\tenpoint{\def\rm{\fam0\tenrm}
  \textfont0=\tenrm \scriptfont0=\sevenrm \scriptscriptfont0=\fiverm
  \textfont1=\teni \scriptfont1=\seveni \scriptscriptfont1=\fivei
  \textfont2=\tensy \scriptfont2=\sevensy \scriptscriptfont2=\fivesy
  \textfont3=\tenex \scriptfont3=\tenex \scriptscriptfont3=\tenex
  \def\mcal{\fam2 \tensy}  \def\mmit{\fam1 \teni}
  \textfont\itfam=\tenit \def\it{\fam\itfam\tenit}
  \textfont\slfam=\tensl \def\sl{\fam\slfam\tensl}
  \textfont\ttfam=\tentt \scriptfont\ttfam=\eighttt
  \scriptscriptfont\ttfam=\eighttt  \def\tt{\fam\ttfam\tentt}
  \textfont\bffam=\tenbf \scriptfont\bffam=\sevenbf
  \scriptscriptfont\bffam=\fivebf \def\bf{\fam\bffam\tenbf}
     \ifx\arisposta\amsrisposta    \ifnum\contaeuler=1
  \textfont\eufmfam=\teneufm \scriptfont\eufmfam=\seveneufm
  \scriptscriptfont\eufmfam=\fiveeufm \def\eufm{\fam\eufmfam\teneufm}
  \textfont\eufbfam=\teneufb \scriptfont\eufbfam=\seveneufb
  \scriptscriptfont\eufbfam=\fiveeufb \def\eufb{\fam\eufbfam\teneufb}
  \def\eurm{\teneurm} \def\eurb{\teneurb} \def\eusm{\teneusm}
  \def\eusb{\teneusb}    \fi    \ifnum\contaams=1
  \textfont\msamfam=\tenmsam \scriptfont\msamfam=\sevenmsam
  \scriptscriptfont\msamfam=\fivemsam \def\msam{\fam\msamfam\tenmsam}
  \textfont\msbmfam=\tenmsbm \scriptfont\msbmfam=\sevenmsbm
  \scriptscriptfont\msbmfam=\fivemsbm \def\msbm{\fam\msbmfam\tenmsbm}
     \fi      \ifnum\contacyrill=1     \def\cyrill{\tenwncyr}
  \def\cyrilb{\tenwncyb}  \def\cyrili{\tenwncyi}         \fi
  \textfont3=\tenex \scriptfont3=\sevenex \scriptscriptfont3=\sevenex
  \def\cmmib{\fam\cmmibfam\tencmmib} \scriptfont\cmmibfam=\sevencmmib
  \textfont\cmmibfam=\tencmmib  \scriptscriptfont\cmmibfam=\fivecmmib
  \def\cmbsy{\fam\cmbsyfam\tencmbsy} \scriptfont\cmbsyfam=\sevencmbsy
  \textfont\cmbsyfam=\tencmbsy  \scriptscriptfont\cmbsyfam=\fivecmbsy
  \def\cmcsc{\fam\cmcscfam\tencmcsc} \scriptfont\cmcscfam=\eightcmcsc
  \textfont\cmcscfam=\tencmcsc \scriptscriptfont\cmcscfam=\eightcmcsc
     \fi            \tt \ttglue=.5em plus.25em minus.15em
  \normalbaselineskip=12pt
  \setbox\strutbox=\hbox{\vrule height8.5pt depth3.5pt width0pt}
  \let\sc=\eightrm \let\big=\tenbig   \normalbaselines
  \baselineskip=\infralinea  \rm}
\gdef\ninepoint{\def\rm{\fam0\ninerm}
  \textfont0=\ninerm \scriptfont0=\sixrm \scriptscriptfont0=\fiverm
  \textfont1=\ninei \scriptfont1=\sixi \scriptscriptfont1=\fivei
  \textfont2=\ninesy \scriptfont2=\sixsy \scriptscriptfont2=\fivesy
  \textfont3=\tenex \scriptfont3=\tenex \scriptscriptfont3=\tenex
  \def\mcal{\fam2 \ninesy}  \def\mmit{\fam1 \ninei}
  \textfont\itfam=\nineit \def\it{\fam\itfam\nineit}
  \textfont\slfam=\ninesl \def\sl{\fam\slfam\ninesl}
  \textfont\ttfam=\ninett \scriptfont\ttfam=\eighttt
  \scriptscriptfont\ttfam=\eighttt \def\tt{\fam\ttfam\ninett}
  \textfont\bffam=\ninebf \scriptfont\bffam=\sixbf
  \scriptscriptfont\bffam=\fivebf \def\bf{\fam\bffam\ninebf}
     \ifx\arisposta\amsrisposta  \ifnum\contaeuler=1
  \textfont\eufmfam=\nineeufm \scriptfont\eufmfam=\sixeufm
  \scriptscriptfont\eufmfam=\fiveeufm \def\eufm{\fam\eufmfam\nineeufm}
  \textfont\eufbfam=\nineeufb \scriptfont\eufbfam=\sixeufb
  \scriptscriptfont\eufbfam=\fiveeufb \def\eufb{\fam\eufbfam\nineeufb}
  \def\eurm{\nineeurm} \def\eurb{\nineeurb} \def\eusm{\nineeusm}
  \def\eusb{\nineeusb}     \fi   \ifnum\contaams=1
  \textfont\msamfam=\ninemsam \scriptfont\msamfam=\sixmsam
  \scriptscriptfont\msamfam=\fivemsam \def\msam{\fam\msamfam\ninemsam}
  \textfont\msbmfam=\ninemsbm \scriptfont\msbmfam=\sixmsbm
  \scriptscriptfont\msbmfam=\fivemsbm \def\msbm{\fam\msbmfam\ninemsbm}
     \fi       \ifnum\contacyrill=1     \def\cyrill{\ninewncyr}
  \def\cyrilb{\ninewncyb}  \def\cyrili{\ninewncyi}         \fi
  \textfont3=\nineex \scriptfont3=\sevenex \scriptscriptfont3=\sevenex
  \def\cmmib{\fam\cmmibfam\ninecmmib}  \textfont\cmmibfam=\ninecmmib
  \scriptfont\cmmibfam=\sixcmmib \scriptscriptfont\cmmibfam=\fivecmmib
  \def\cmbsy{\fam\cmbsyfam\ninecmbsy}  \textfont\cmbsyfam=\ninecmbsy
  \scriptfont\cmbsyfam=\sixcmbsy \scriptscriptfont\cmbsyfam=\fivecmbsy
  \def\cmcsc{\fam\cmcscfam\ninecmcsc} \scriptfont\cmcscfam=\eightcmcsc
  \textfont\cmcscfam=\ninecmcsc \scriptscriptfont\cmcscfam=\eightcmcsc
     \fi            \tt \ttglue=.5em plus.25em minus.15em
  \normalbaselineskip=11pt
  \setbox\strutbox=\hbox{\vrule height8pt depth3pt width0pt}
  \let\sc=\sevenrm \let\big=\ninebig \normalbaselines\rm}
\gdef\eightpoint{\def\rm{\fam0\eightrm}
  \textfont0=\eightrm \scriptfont0=\sixrm \scriptscriptfont0=\fiverm
  \textfont1=\eighti \scriptfont1=\sixi \scriptscriptfont1=\fivei
  \textfont2=\eightsy \scriptfont2=\sixsy \scriptscriptfont2=\fivesy
  \textfont3=\tenex \scriptfont3=\tenex \scriptscriptfont3=\tenex
  \def\mcal{\fam2 \eightsy}  \def\mmit{\fam1 \eighti}
  \textfont\itfam=\eightit \def\it{\fam\itfam\eightit}
  \textfont\slfam=\eightsl \def\sl{\fam\slfam\eightsl}
  \textfont\ttfam=\eighttt \scriptfont\ttfam=\eighttt
  \scriptscriptfont\ttfam=\eighttt \def\tt{\fam\ttfam\eighttt}
  \textfont\bffam=\eightbf \scriptfont\bffam=\sixbf
  \scriptscriptfont\bffam=\fivebf \def\bf{\fam\bffam\eightbf}
     \ifx\arisposta\amsrisposta   \ifnum\contaeuler=1
  \textfont\eufmfam=\eighteufm \scriptfont\eufmfam=\sixeufm
  \scriptscriptfont\eufmfam=\fiveeufm \def\eufm{\fam\eufmfam\eighteufm}
  \textfont\eufbfam=\eighteufb \scriptfont\eufbfam=\sixeufb
  \scriptscriptfont\eufbfam=\fiveeufb \def\eufb{\fam\eufbfam\eighteufb}
  \def\eurm{\eighteurm} \def\eurb{\eighteurb} \def\eusm{\eighteusm}
  \def\eusb{\eighteusb}       \fi    \ifnum\contaams=1
  \textfont\msamfam=\eightmsam \scriptfont\msamfam=\sixmsam
  \scriptscriptfont\msamfam=\fivemsam \def\msam{\fam\msamfam\eightmsam}
  \textfont\msbmfam=\eightmsbm \scriptfont\msbmfam=\sixmsbm
  \scriptscriptfont\msbmfam=\fivemsbm \def\msbm{\fam\msbmfam\eightmsbm}
     \fi       \ifnum\contacyrill=1     \def\cyrill{\eightwncyr}
  \def\cyrilb{\eightwncyb}  \def\cyrili{\eightwncyi}         \fi
  \textfont3=\eightex \scriptfont3=\sevenex \scriptscriptfont3=\sevenex
  \def\cmmib{\fam\cmmibfam\eightcmmib}  \textfont\cmmibfam=\eightcmmib
  \scriptfont\cmmibfam=\sixcmmib \scriptscriptfont\cmmibfam=\fivecmmib
  \def\cmbsy{\fam\cmbsyfam\eightcmbsy}  \textfont\cmbsyfam=\eightcmbsy
  \scriptfont\cmbsyfam=\sixcmbsy \scriptscriptfont\cmbsyfam=\fivecmbsy
  \def\cmcsc{\fam\cmcscfam\eightcmcsc} \scriptfont\cmcscfam=\eightcmcsc
  \textfont\cmcscfam=\eightcmcsc \scriptscriptfont\cmcscfam=\eightcmcsc
     \fi             \tt \ttglue=.5em plus.25em minus.15em
  \normalbaselineskip=9pt
  \setbox\strutbox=\hbox{\vrule height7pt depth2pt width0pt}
  \let\sc=\sixrm \let\big=\eightbig \normalbaselines\rm }
\gdef\tenbig#1{{\hbox{$\left#1\vbox to8.5pt{}\right.\n@space$}}}
\gdef\ninebig#1{{\hbox{$\textfont0=\tenrm\textfont2=\tensy
   \left#1\vbox to7.25pt{}\right.\n@space$}}}
\gdef\eightbig#1{{\hbox{$\textfont0=\ninerm\textfont2=\ninesy
   \left#1\vbox to6.5pt{}\right.\n@space$}}}
\def\alternativefont#1#2{\ifx\arisposta\amsrisposta \relax \else
\xdef#1{#2} \fi}
\global\contaeuler=0 \global\contacyrill=0 \global\contaams=0
%
%
%
%
\newbox\fotlinebb \newbox\hedlinebb \newbox\leftcolumn
\gdef\makeheadline{\vbox to 0pt{\vskip-22.5pt
     \fullline{\vbox to8.5pt{}\the\headline}\vss}\nointerlineskip}
\gdef\makehedlinebb{\vbox to 0pt{\vskip-22.5pt
     \fullline{\vbox to8.5pt{}\copy\hedlinebb\hfil
     \line{\hfill\the\headline\hfill}}\vss} \nointerlineskip}
\gdef\makefootline{\baselineskip=24pt \fullline{\the\footline}}
\gdef\makefotlinebb{\baselineskip=24pt
    \fullline{\copy\fotlinebb\hfil\line{\hfill\the\footline\hfill}}}
\gdef\doubleformat{\shipout\vbox{\Landspec\makehedlinebb
     \fullline{\box\leftcolumn\hfil\columnbox}\makefotlinebb}
     \advancepageno}
\gdef\columnbox{\leftline{\pagebody}}
\gdef\line#1{\hbox to\hsize{\hskip\leftskip#1\hskip\rightskip}}
\gdef\fullline#1{\hbox to\fullhsize{\hskip\leftskip{#1}%
\hskip\rightskip}}
\gdef\footnote#1{\let\@sf=\empty
         \ifhmode\edef\#sf{\spacefactor=\the\spacefactor}\/\fi
         #1\@sf\vfootnote{#1}}
\gdef\vfootnote#1{\insert\footins\bgroup
         \ifnum\dimnota=1  \eightpoint\fi
         \ifnum\dimnota=2  \ninepoint\fi
         \ifnum\dimnota=0  \tenpoint\fi
         \interlinepenalty=\interfootnotelinepenalty
         \splittopskip=\ht\strutbox
         \splitmaxdepth=\dp\strutbox \floatingpenalty=20000
         \leftskip=\oldssposta \rightskip=\olddsposta
         \spaceskip=0pt \xspaceskip=0pt
         \ifnum\sinnota=0   \textindent{#1}\fi
         \ifnum\sinnota=1   \item{#1}\fi
         \footstrut\futurelet\next\fo@t}
\gdef\fo@t{\ifcat\bgroup\noexpand\next \let\next\f@@t
             \else\let\next\f@t\fi \next}
\gdef\f@@t{\bgroup\aftergroup\@foot\let\next}
\gdef\f@t#1{#1\@foot} \gdef\@foot{\strut\egroup}
\gdef\footstrut{\vbox to\splittopskip{}}
\skip\footins=\bigskipamount
\count\footins=1000  \dimen\footins=8in
\catcode`@=12
\tenpoint
\ifnum\unoduecol=1 \hsize=\tothsize   \fullhsize=\tothsize \fi
\ifnum\unoduecol=2 \hsize=\collhsize  \fullhsize=\tothsize \fi
\global\let\lrcol=L      \ifnum\unoduecol=1
\output{\plainoutput{\ifnum\tipbnota=2 \clearnmbnota\fi}} \fi
\ifnum\unoduecol=2 \output{\if L\lrcol
     \global\setbox\leftcolumn=\columnbox
     \global\setbox\fotlinebb=\line{\hfill\the\footline\hfill}
     \global\setbox\hedlinebb=\line{\hfill\the\headline\hfill}
     \advancepageno  \global\let\lrcol=R
     \else  \doubleformat \global\let\lrcol=L \fi
     \ifnum\outputpenalty>-20000 \else\dosupereject\fi
     \ifnum\tipbnota=2\clearnmbnota\fi }\fi
\def\ifdoublepage{\ifnum\unoduecol=2 }
\gdef\yespagenumbers{\footline={\hss\tenrm\folio\hss}}
\gdef\ciao{ \ifnum\fdefcontre=1 \endfdef\fi
     \par\vfill\supereject \ifnum\unoduecol=2
     \if R\lrcol  \headline={}\nopagenumbers\null\vfill\eject
     \fi\fi \end}

\newskip\olddsposta \newskip\oldssposta
\global\oldssposta=\leftskip \global\olddsposta=\rightskip

\def\filldots{\leaders\hbox to 1em{\hss.\hss}\hfill}
\def\inquadrb#1 {\vbox {\hrule  \hbox{\vrule \vbox {\vskip .2cm
    \hbox {\ #1\ } \vskip .2cm } \vrule  }  \hrule} }
 \def\newline{\hfil\break}
\def\jump{\vskip\baselineskip} \newskip\iinnffrr
\def\sjump{\iinnffrr=\baselineskip
          \divide\iinnffrr by 2 \vskip\iinnffrr}
\def\bjump{\vskip\baselineskip \vskip\baselineskip}
\newcount\nmbnota  \def\clearnmbnota{\global\nmbnota=0}
\newcount\tipbnota \def\letterfootnote{\global\tipbnota=1}

\def\note#1{\global\advance\nmbnota by 1 \ifnum\tipbnota=1
    \footnote{$^{\rm\nttlett}$}{#1} \else {\ifnum\tipbnota=2
    \footnote{$^{\nttsymb}$}{#1}
    \else\footnote{$^{\the\nmbnota}$}{#1}\fi}\fi}
\def\nttlett{\ifcase\nmbnota \or a\or b\or c\or d\or e\or f\or
g\or h\or i\or j\or k\or l\or m\or n\or o\or p\or q\or r\or
s\or t\or u\or v\or w\or y\or x\or z\fi}
\def\nttsymb{\ifcase\nmbnota \or\dag\or\sharp\or\ddag\or\star\or
\natural\or\flat\or\clubsuit\or\diamondsuit\or\heartsuit
\or\spadesuit\fi}   \clearnmbnota
\def\numberfootnote{\global\tipbnota=0} \numberfootnote
\def\setnote#1{\expandafter\xdef\csname#1\endcsname{
\ifnum\tipbnota=1 {\rm\nttlett} \else {\ifnum\tipbnota=2
{\nttsymb} \else \the\nmbnota\fi}\fi} }
\newcount\nbmfig  \def\clearnbmfig{\global\nbmfig=0}
\gdef\figure{\global\advance\nbmfig by 1
      {\rm fig. \the\nbmfig}}   \clearnbmfig
\def\setfig#1{\expandafter\xdef\csname#1\endcsname{fig. \the\nbmfig}}
 \def\endformula{\eqno\numero $$}
 \def\efr{\endformula}
\newcount\frmcount \def\clearfrmcount{\global\frmcount=0}
\def\numero{\global\advance\frmcount by 1   \ifnum\indappcount=0
  {\ifnum\cpcount <1 {\hbox{\rm (\the\frmcount )}}  \else
  {\hbox{\rm (\the\cpcount .\the\frmcount )}} \fi}  \else
  {\hbox{\rm (\applett .\the\frmcount )}} \fi}
\def\nameformula#1{\global\advance\frmcount by 1%
\ifnum\draftnum=0  {\ifnum\indappcount=0%
{\ifnum\cpcount<1\xdef\spzzttrra{(\the\frmcount )}%
\else\xdef\spzzttrra{(\the\cpcount .\the\frmcount )}\fi}%
\else\xdef\spzzttrra{(\applett .\the\frmcount )}\fi}%
\else\xdef\spzzttrra{(#1)}\fi%
\expandafter\xdef\csname#1\endcsname{\spzzttrra}
\eqno \hbox{\rm\spzzttrra} $$}
\def\nfr{\nameformula}    
\def\nameali#1{\global\advance\frmcount by 1%
\ifnum\draftnum=0  {\ifnum\indappcount=0%
{\ifnum\cpcount<1\xdef\spzzttrra{(\the\frmcount )}%
\else\xdef\spzzttrra{(\the\cpcount .\the\frmcount )}\fi}%
\else\xdef\spzzttrra{(\applett .\the\frmcount )}\fi}%
\else\xdef\spzzttrra{(#1)}\fi%
\expandafter\xdef\csname#1\endcsname{\spzzttrra}
  \hbox{\rm\spzzttrra} }      \clearfrmcount
\newcount\cpcount \def\clearcpcount{\global\cpcount=0}
\newcount\subcpcount \def\clearsubcpcount{\global\subcpcount=0}
\newcount\appcount \def\clearappcount{\global\appcount=0}
\newcount\indappcount \def\clearindappcount{\indappcount=0}
\newcount\sottoparcount 

\def\applett{\ifcase\appcount  \or {A}\or {B}\or {C}\or
{D}\or {E}\or {F}\or {G}\or {H}\or {I}\or {J}\or {K}\or {L}\or
{M}\or {N}\or {O}\or {P}\or {Q}\or {R}\or {S}\or {T}\or {U}\or
{V}\or {W}\or {X}\or {Y}\or {Z}\fi    \ifnum\appcount<0
\immediate\write16 {Panda ERROR - Appendix: counter "appcount"
out of range}\fi  \ifnum\appcount>26  \immediate\write16 {Panda
ERROR - Appendix: counter "appcount" out of range}\fi}
\clearappcount  \clearindappcount \newcount\connttrre
\def\clearconnttrre{\global\connttrre=0} \newcount\countref
\def\clearcountref{\global\countref=0} \clearcountref
\def\chapter#1{\global\advance\cpcount by 1 \clearfrmcount
                 \goodbreak\null\vbox{\jump\nobreak
                 \clearsubcpcount\clearindappcount
                 \itemitem{\ttaarr\the\cpcount .\qquad}{\ttaarr #1}
                 \par\nobreak\jump\sjump}\nobreak}
\def\section#1{\global\advance\subcpcount by 1 \goodbreak\null
               \vbox{\sjump\nobreak\ifnum\indappcount=0
                 {\ifnum\cpcount=0 {\itemitem{\ppaarr
               .\the\subcpcount\quad\enskip\ }{\ppaarr #1}\par} \else
                 {\itemitem{\ppaarr\the\cpcount .\the\subcpcount\quad
                  \enskip\ }{\ppaarr #1} \par}  \fi}
                \else{\itemitem{\ppaarr\applett .\the\subcpcount\quad
                 \enskip\ }{\ppaarr #1}\par}\fi\nobreak\jump}\nobreak}
\clearsubcpcount
\def\appendix#1{\global\advance\appcount by 1 \clearfrmcount
                  \goodbreak\null\vbox{\jump\nobreak
                  \global\advance\indappcount by 1 \clearsubcpcount
          \itemitem{ }{\hskip-40pt\ttaarr Appendix\ \applett :\ #1}
             \nobreak\jump\sjump}\nobreak}
\clearappcount \clearindappcount
\def\references{\goodbreak\null\vbox{\jump\nobreak
   \itemitem{}{\ttaarr References} \nobreak\jump\sjump}\nobreak}

\clearcpcount\clearcountref

\def\setchap#1{\ifnum\indappcount=0{\ifnum\subcpcount=0%
\xdef\spzzttrra{\the\cpcount}%
\else\xdef\spzzttrra{\the\cpcount .\the\subcpcount}\fi}
\else{\ifnum\subcpcount=0 \xdef\spzzttrra{\applett}%
\else\xdef\spzzttrra{\applett .\the\subcpcount}\fi}\fi
\expandafter\xdef\csname#1\endcsname{\spzzttrra}}
\newcount\draftnum \newcount\ppora   \newcount\ppminuti
\global\ppora=\time   \global\ppminuti=\time
\global\divide\ppora by 60  \draftnum=\ppora
\multiply\draftnum by 60    \global\advance\ppminuti by -\draftnum
\def\droggi{\number\day /\number\month /\number\year\ \the\ppora
:\the\ppminuti}     \global\draftnum=0
\def\draftcomment#1{\ifnum\draftnum=0 \relax \else
{\ {\bf ***}\ #1\ {\bf ***}\ }\fi} 
%
%
\catcode`@=11
\gdef\Ref#1{\expandafter\ifx\csname @rrxx@#1\endcsname\relax%
{\global\advance\countref by 1    \ifnum\countref>200
\immediate\write16 {Panda ERROR - Ref: maximum number of references
exceeded}  \expandafter\xdef\csname @rrxx@#1\endcsname{0}\else
\expandafter\xdef\csname @rrxx@#1\endcsname{\the\countref}\fi}\fi
\ifnum\draftnum=0 \csname @rrxx@#1\endcsname \else#1\fi}
\gdef\beginref{\ifnum\draftnum=0  \gdef\Rref{\fairef}
\gdef\endref{\scriviref} \else\relax\fi
\ifx\risposta\mplarisposta \ninepoint \fi
\parskip 2pt plus.2pt \baselineskip=12pt}
\def\Reflab#1{[#1]} \gdef\Rref#1#2{\item{\Reflab{#1}}{#2}}
\gdef\endref{\relax}  \newcount\conttemp
\gdef\fairef#1#2{\expandafter\ifx\csname @rrxx@#1\endcsname\relax
{\global\conttemp=0 \immediate\write16 {Panda ERROR - Ref: reference
[#1] undefined}} \else
{\global\conttemp=\csname @rrxx@#1\endcsname } \fi
\global\advance\conttemp by 50  \global\setbox\conttemp=\hbox{#2} }
\gdef\scriviref{\clearconnttrre\conttemp=50
\loop\ifnum\connttrre<\countref \advance\conttemp by 1
\advance\connttrre by 1
\item{\Reflab{\the\connttrre}}{\unhcopy\conttemp} \repeat}
\clearcountref \clearconnttrre
\catcode`@=12
\ifx\risposta\mplarisposta \def\Reflab#1{#1.} \letterfootnote \fi

\def\slashchar#1{\setbox0=\hbox{$#1$} \dimen0=\wd0
     \setbox1=\hbox{/} \dimen1=\wd1 \ifdim\dimen0>\dimen1
      \rlap{\hbox to \dimen0{\hfil/\hfil}} #1 \else
      \rlap{\hbox to \dimen1{\hfil$#1$\hfil}} / \fi}
\ifx\oldchi\undefined \let\oldchi=\chi
  \def\cchi{{\raise 1pt\hbox{$\oldchi$}}} \let\chi=\cchi \fi

\def\frac#1#2{{\textstyle{#1 \over #2}}}

\def\half{\ifinner {\scriptstyle {1 \over 2}}\else {1 \over 2} \fi}

\def\simge{\rlap{\raise 2pt \hbox{$>$}}{\lower 2pt \hbox{$\sim$}}}
\def\simle{\rlap{\raise 2pt \hbox{$<$}}{\lower 2pt \hbox{$\sim$}}}

\def\vbig#1#2{{\vbigd@men=#2\divide\vbigd@men by 2%
\hbox{$\left#1\vbox to \vbigd@men{}\right.\n@space$}}}

%
%
\newcount\fdefcontre \newcount\fdefcount \newcount\indcount
\newread\filefdef  \newread\fileftmp  \newwrite\filefdef
\newwrite\fileftmp     \def\strip#1*.A {#1}
\def\futuredef#1{\beginfdef
\expandafter\ifx\csname#1\endcsname\relax%
{\immediate\write\fileftmp {#1*.A}
\immediate\write16 {Panda Warning - fdef: macro "#1" on page
\the\pageno \space undefined}
\ifnum\draftnum=0 \expandafter\xdef\csname#1\endcsname{(?)}
\else \expandafter\xdef\csname#1\endcsname{(#1)} \fi
\global\advance\fdefcount by 1}\fi   \csname#1\endcsname}

\def\beginfdef{\ifnum\fdefcontre=0
\immediate\openin\filefdef \jobname.fdef
\immediate\openout\fileftmp \jobname.ftmp
\global\fdefcontre=1  \ifeof\filefdef \immediate\write16 {Panda
WARNING - fdef: file \jobname.fdef not found, run TeX again}
\else \immediate\read\filefdef to\spzzttrra
\global\advance\fdefcount by \spzzttrra
\indcount=0      \loop\ifnum\indcount<\fdefcount
\advance\indcount by 1   \immediate\read\filefdef to\spezttrra
\immediate\read\filefdef to\sppzttrra
\edef\spzzttrra{\expandafter\strip\spezttrra}
\immediate\write\fileftmp {\spzzttrra *.A}
\expandafter\xdef\csname\spzzttrra\endcsname{\sppzttrra}
\repeat \fi \immediate\closein\filefdef \fi}
\def\endfdef{\immediate\closeout\fileftmp   \ifnum\fdefcount>0
\immediate\openin\fileftmp \jobname.ftmp
\immediate\openout\filefdef \jobname.fdef
\immediate\write\filefdef {\the\fdefcount}   \indcount=0
\loop\ifnum\indcount<\fdefcount    \advance\indcount by 1
\immediate\read\fileftmp to\spezttrra
\edef\spzzttrra{\expandafter\strip\spezttrra}
\immediate\write\filefdef{\spzzttrra *.A}
\edef\spezttrra{\string{\csname\spzzttrra\endcsname\string}}
\iwritel\filefdef{\spezttrra}
\repeat  \immediate\closein\fileftmp \immediate\closeout\filefdef
\immediate\write16 {Panda Warning - fdef: Label(s) may have changed,
re-run TeX to get them right}\fi}
\def\iwritel#1#2{\newlinechar=-1
{\newlinechar=`\ \immediate\write#1{#2}}\newlinechar=-1}
\global\fdefcontre=0 \global\fdefcount=0 \global\indcount=0
%
%
\null
%
%
%
%

%
\loadamsmath
%
\pageno=0\baselineskip=14pt
\nopagenumbers{
\line{\hfill CERN-TH.7164/94}
\line{\hfill SWAT/93-94/26}
\line{\hfill\tt hep-th/9402084}
\line{\hfill February 1994}
\ifdoublepage \bjump\bjump\bjump\bjump\else\vfill\fi
\centerline{\capsone The exact mass-gaps of the principal chiral
models}
\bjump\bjump
\centerline{\scaps Timothy J. Hollowood\footnote{$^*$}{On leave from:
Department of Mathematics, University College of Swansea, Swansea, SA2
8PP, U.K.}}
\sjump
\sjump
\centerline{\sl CERN-TH, CH-1211 Geneva 23, Switzerland.}
\centerline{\tt hollow@surya11.cern.ch}
\bjump\bjump\bjump
\ifdoublepage
\vfill
\noindent
\line{CERN-TH.7164/94\hfill}
\line{February 1994\hfill}
\eject\null\vfill\fi
\centerline{\capsone ABSTRACT}\sjump
An exact expression for the mass-gap, the ratio of the physical
particle mass to the $\Lambda$-parameter, is found for the principal
chiral sigma models associated to all the classical Lie algebras. The
calculation is based on a comparison of the free-energy in the presence
of a source coupling to a conserved charge of the theory computed in
two ways: via the thermodynamic Bethe Ansatz from the exact
scattering matrix and directly in perturbation theory. The calculation
provides a non-trivial test of the form of the exact scattering matrix.
\sjump\vfill
\ifdoublepage \else
\noindent
\line{CERN-TH.7164/94\hfill}
\line{February 1994\hfill}\fi
\eject}
\yespagenumbers\pageno=1
%
%
\def\t{\theta}

\chapter{Introduction}

The principal chiral, or ${\cal G}\times{\cal G}$, sigma models are
two-dimensional quantum field
theories that are integrable at the quantum level. The
fact that the theories are integrable means that their scattering
matrices are factorizable. Such $S$-matrices have been
conjectured for all the theories corresponding to the classical Lie
algebras [\Ref{ORW}]. Expressions for the complete $S$-matrices for
SU($N$) can be found in [\Ref{ORW}] and for Sp($2N$) in [\Ref{TH}].
Not all the $S$-matrix elements are known in the case of
SO($N$). It is an outstanding problem to prove
from first principles that the $S$-matrices actually do describe the
lagrangian field theories. This is
especially important because a given factorizable $S$-matrix is
ambiguous since it may
always by multiplied by CDD factors. Connecting the $S$-matrix picture
with the lagrangian picture is highly non-trivial since
the masses are generated dynamically and the theories are
asymptotically free.

In a series of papers such non-trivial tests have been applied
to various integrable models: the Gross-Neveu model [\Ref{FNW}], the O($N$)
sigma model [\Ref{HMN},\Ref{HN}] and the SU($N$) principal chiral sigma model
[\Ref{BNNW},\Ref{W}] (the SU(2) case was also considered in
[\Ref{PW},\Ref{Y}]),
using a technique known as the Thermodynamic Bethe Ansatz
(TBA). The central idea is to couple the theory to a particular
conserved current
and then compute the response of the free-energy for large
values of the source in the regime when
conventional perturbation theory is valid.
The same quantity can then be computed directly from the $S$-matrix
using the TBA equations at zero temperature in which the coupling to the source
appears as a chemical potential. By comparing the two
expressions a non-trivial test of the $S$-matrix is obtained as well
as an exact expression for the mass-gap (the ratio of the physical
mass to the $\Lambda$-parameter).

In general the solution of the TBA equations at zero temperature
coupled to an arbitrary chemical potential would be a formidable
problem, even in ultra-violet limit, since the equations are a set of
coupled integral equations. However, by a judicious choice of the
source the state of the system contains just one particle which
undergoes elastic scattering
(the particle is the highest weight state of
a multiplet). The TBA equations then reduce to a single integral
equation which can be solved in the ultra-violet limit using
generalized Wiener-Hopf techniques [\Ref{HMN},\Ref{JNW}]
(for a summary see the appendix of [\Ref{FNW}]).

In this paper we extend the results of [\Ref{BNNW}] to the principal chiral
models for all the
classical Lie algebras and arrive at a universal formula for the exact
mass-gap. We also show that by tuning the
source we can force the system into inequivalent ground-states which
each consist of a single type of particle. The fact that the
ground-states are pure for particular values of the source, is presented
as a conjecture whose ultimate justification comes from the agreement
with perturbation theory; however, it should be possible to prove this
fact directly from the full TBA equations of the models.

The principal chiral models are described by a
lagrangian density
$$
{\cal L}_0=-{1\over\lambda^2}{\rm Tr}\left(g^{-1}\partial_\mu g\cdot
g^{-1}\partial^\mu g\right),
\efr
where $g$ is a group valued field. The theory is invariant under a
global symmetry corresponding to left and right multiplication by the
group $g\mapsto h_{\rm L}gh_{\rm R}^{-1}$, $h_{\rm L,R}\in{\cal G}$.
$\lambda$ is a dimensionless coupling constant.

The $S$-matrices that have been conjectured to describe the scattering
of the states of the model describe $r={\rm rank}({\cal G})$
particles which are associated to the fundamental representations of
${\cal G}$. The masses of the particles, except for those associated to
the spinors of SO($N$), can be described by the
universal formula [\Ref{ORW}]
$$
m_a=m{\sin(\pi a/g)\over\sin(\pi/g)},\qquad a=1,2,\ldots,
\nfr{MASS}
where $g$ is the dual Coxeter number of the Lie
algebra associated to the group: for
${\rm A}_r$, ${\rm B}_r$, ${\rm C}_r$ and ${\rm D}_r$ it is $r+1$,
$2r-1$, $2(r+1)$ and $2(r-1)$,
respectively. The masses of the spinors
of ${\rm B}_r$ and ${\rm D}_r$ are
$$
{\rm B}_r:\quad m_r={m\over2\sin(\pi/g)},\quad {\rm D}_r:\quad
m_{r-1}=m_r={m\over2\sin(\pi/g)}.
\efr
The particle with mass $m_a$ transforms in the following representation
of ${\cal G}\times{\cal G}$ [\Ref{ORW}]:
$$\eqalign{
{\rm A}_r:\qquad &W_a=V_a\otimes V_a,\quad
a=1,2,\ldots,r,\cr
{\rm B}_r:\qquad
&W_a=\sum_{k=0}^{a-2k\geq0}V_{a-2k}\otimes\sum_{j=0}^{a-2j\geq0}V_{a-2j},\quad
a=1,2,\ldots,r-1,\cr
&W_r=V_r\otimes V_r,\cr
{\rm C}_r:\qquad &W_a=V_a\otimes V_a,\quad
a=1,2,\ldots,r,\cr
{\rm D}_r:\qquad &W_a=\sum_{k=0}^{a-2k\geq0}V_{a-2k}\otimes
\sum_{j=0}^{a-2j\geq0}V_{a-2j},\quad
a=1,2,\ldots,r-2,\cr
&W_{r-1}=V_{r-1}\otimes V_{r-1},\qquad W_r=V_r\otimes V_r,\cr}
\efr
where $V_a$ is the $a^{\rm th}$ fundamental representation of ${\cal
G}$ with the standard labelling of the Dynkin diagram [\Ref{ORW}].
Notice that although the
particles are associated to the fundamental representations
they are sometimes reducible in the case of SO($N$).

Fortunately, we shall not require the expression for the complete
$S$-matrices but only those elements for the
particles of the highest weight in
each multiplet (so with quantum numbers $|\omega_a,\omega_a\rangle$
where the $\omega_a$'s are the fundamental weights). The $S$-matrix
amongst these states is purely elastic and their expressions can
be extracted from [\Ref{ORW}]:
$$
S_{ab}(\t)=\exp\left\{i\pi\delta_{ab}+
2i\int_0^\infty{dx\over x}\sin(\t x)\left[R_{ab}
(x)-\delta_{ab}\right]\right\},
\nfr{SM}
where $\t$ is the rapidity difference of the incoming particles and the
kernel $R_{ab}(\t)$ has the following form for all the particles
except the spinors:
$$\eqalign{
{\rm A}_r:\qquad&R_{ab}(x)={2\sinh\left({{\rm min}(a,b)\over r+1}\pi x\right)
\sinh\left({r+1-{\rm max}(a,b)\over r+1}\pi x\right)\over\sinh\left({1\over
r+1}\pi x\right)},\cr
{\rm B}_r,{\rm C}_r,{\rm D}_r:\qquad&R_{ab}(x)={2\sinh\left(
{{\rm min}(a,b)\over g}\pi x\right)\cosh\left(
{g-2{\rm max}(a,b)\over2g}\pi x\right)\over\cosh\left({1\over2}\pi
x\right)}.\cr}
\efr
The $S$-matrix elements involving the spinors can also be deduced from the
formulas of [\Ref{ORW}] but we shall not require them.

In the following two section we calculate the free-energy in the
presence of a source coupling to the conserved charge of the ${\cal
G}\times{\cal G}$ symmetry in two ways: from the
lagrangian using perturbation theory and from the $S$-matrix using the
thermodynamic Bethe Ansatz.

\chapter{Free-energy in perturbation theory}

The conserved currents of the left and right symmetry are $J^{\rm
L}_\mu=g^{-1}\partial_\mu g$ and $J^{\rm R}_\mu=(\partial_\mu g)g^{-1}$. We
wish to couple to modify the hamiltonian of the theory by introducing a
coupling to the conserved charge of the diagonal action of the
symmetry. At the lagrangian level this is described by introducing the
``covariant derivative'' [\Ref{BNNW}]:
$$
D_\mu g=\partial_\mu g-ih\delta_{\mu0}(Qg+gQ),
\efr
where $Q$ is a constant element of the Lie algebra. The lagrangian
density in the presence of the source is
$$
{\cal L}={\cal L}_0-{2hi\over\lambda^2}{\rm
Tr}\left((g^{-1}Q+Qg^{-1})\partial_0g\right)-{2h^2\over\lambda^2}{\rm
Tr}\left(Q^2+g^{-1}QgQ\right).
\nfr{LAGS}

The quantity we will calculate is $\delta f(h)=f(h)-f(0)$ where
$f(h)$ is the free-energy per unit volume in the presence of the
source. We shall perform a perturbative calculation in the running
coupling $\lambda(h)$ which in the ultra-violet regime (large $h$)
runs to zero and hence is the regime where perturbation theory will
be reliable. We shall only perform the computation to one loop;
however this will be sufficient to provide a non-trivial check of the
$S$-matrix and allow for the evaluation of the exact mass-gap.

An explicit basis for $g$ is provided by
$$
g=\exp\left\{i\sum_{\alpha}n^{(\alpha)}E_\alpha+in\cdot H\right\},
\efr
where the fields satisfy the reality condition
$n^{(\alpha)*}=n^{(-\alpha)}$ and $n^*=n$, and the sum is
over all the roots of algebra. In the above $E_\alpha$ is the usual
step generator associated to a root $\alpha$ and $H$ is the generator
of the Cartan subalgebra. In what follows we choose a normalization
in which the roots of the simply-laced algebras have length-squared 2
and the the long roots of ${\rm B}_r$ and ${\rm C}_r$
have length-squared $2$ and $4$, respectively.

Without loss of generality we take $Q$ to be in the Cartan subalgebra so
$Q=q\cdot H$, where $q$ is some $r$-dimensional vector.
The quadratic part of the (euclidean) lagrangian density \LAGS\ is
simply
$$
{\cal L}=-{4h^2\over\lambda^2}q^2+
{1\over\lambda^2}\sum_{\alpha>0}\left\{\partial_\mu n^{(\alpha)}
\partial^\mu n^{(-\alpha)}+h^2(\alpha\cdot
q)^2n^{(\alpha)}n^{(-\alpha)}\right\},
\efr
where the sum is over the positive roots and for simplicity we have
changed the normalization of some of the $n^{(\alpha)}$'s.
Notice that the Cartan subalgebra fields $n$ are completely decoupled
to this order in the loop expansion.

The tree level contribution to $\delta f(h)$ is simply
$$
\delta f(h)_0=-{4h^2\over\lambda^2}q^2.
\efr
To evaluate the one-loop contribution we use dimensional
regularization. Using standard methods one finds
$$
\delta f(h)_1=-{h^2g\over2\pi\epsilon}q^2
+{h^2\over4\pi}\sum_{\alpha>0}(\alpha\cdot q)^2\left\{1-\gamma_{\rm
E}+\ln4\pi-\ln\left(h^2(\alpha\cdot q)^2/\mu^2\right)\right\}+\cdots,
\efr
where $\epsilon=d-2$, $\mu$ is the usual mass parameter of dimensional
regularization and $g$ is
the dual Coxeter number as before. To cancel the divergence in the
$\overline{\rm MS}$-scheme we add to the lagrangian a counter-term
$$
\delta{\cal L}=
{h^2g\over2\pi\epsilon}q^2+{h^2g\over4\pi}
q^2(\gamma_{\rm E}-\ln4\pi).
\efr
The quantity $\delta f(h)$ is renormalization group invariant when
$\lambda$ runs with $\mu$. We can use this freedom to set
$\mu=h$. The way that the coupling constant
runs with $h$ is determined from the form of the counter-term. One finds
$$
h{\partial\over\partial h}\lambda^2=-{g\over8\pi}\lambda^4-\beta_1
\lambda^6-{\cal O}(\lambda^8),
\nfr{BFE}
although the second universal coefficient of the beta-function $\beta_1$ is not
determined at the one-loop level. The expression for the
first coefficient of the beta-function $\beta_0=g/8\pi$ agrees with
[\Ref{MS}].

The expression for the free-energy is then
$$
\delta f(h)=-{4h^2\over\lambda^2(h)}q^2-{h^2\over4\pi}\sum_{\alpha>0}
(q\cdot\alpha)^2\left[\ln(q\cdot\alpha)^2-1\right]+{\cal
O}(\lambda^2),
\nfr{HH}
where the explicit $h$ dependence is obtained by expressing
the running coupling in terms of the $\Lambda$-parameter by solving
\BFE:
$$
{1\over\lambda^2(h)}=\beta_0\ln{h\over\Lambda_{\overline{\rm
MS}}}+{\beta_1\over\beta_0}
\ln\ln{h\over\Lambda_{\overline{\rm MS}}}+{\cal
O}\left(1\over\ln{h\over\Lambda_
{\overline{\rm MS}}}\right),
\efr
where $\beta_0=g/8\pi$. Equation \HH\ is the generalization to all the
classical Lie algebras of equation (17) of [\Ref{BNNW}] for ${\rm A}_r$.

For comparing with the expression for the free-energy from the TBA
calculation we set $q=\omega_a/(2\omega_a^2)$
(excluding the spinors of SO($N$)). Writing
$$
\delta f(h)=-{h^2\over4}k^2_a\left[\ln{h\over\Lambda_{\overline{\rm
MS}}}+A_a+{\beta_1\over\beta_0^2}\ln\ln{h\over\Lambda_{\overline{\rm
MS}}}+{\cal O}\left({1\over\ln{h\over\Lambda_{\overline{\rm MS}}}}\right)
\right].
\nfr{PFE}
By explicit computation we find for ${\rm A}_r$ that
$$
k_a^2={(r+1)^2\over 2\pi a(r+1-a)},\qquad
A_a=\ln\left({r+1\over2a(r+1-a)}\right)-{1\over2},
\nfr{KA}
and for the other algebras a universal form applies:
$$
k^2_a={g\over2\pi a},\qquad
A_a=-\ln a-{1\over2}-{d_1-2a\over g}\ln2,
\nfr{KR}
where the quantity $d_1$ is the dimension of the vector representation
of the algebra, i.e. $r+1$, $2r+1$, $2r$ and $2r$ for ${\rm A}_r$,
${\rm B}_r$, ${\rm C}_r$ and ${\rm D}_r$, respectively.

\chapter{Free-energy from the $S$-matrix}

In this section we will calculate $\delta f(h)$ in the ultra-violet
limit, $h\gg m$ directly from the $S$-matrix. The technique is known
as the Thermodynamic Bethe Ansatz (TBA) and in its most general form
it allows one to calculate
the behaviour of the free-energy of a one-dimensional gas of particles
described by a factorizable $S$-matrix on the temperature and in the
presence of a chemical potential. The free-energy is given in terms of
a set of functions---in general infinite in number---which satisfy a set
of coupled integral equations (the TBA equations).

For our application we working on
the plane and hence at zero temperature. The coupling of the theory to
the source in \LAGS\ leads to a particular form for the chemical
potential. The one-particle states are labelled by two
weight vectors $|\mu,\nu\rangle$ (as well as the rapidity) and they
can be chosen to be eigenstates of the charge $Q$ with
$Q|\mu,\nu\rangle=q\cdot(\mu+\nu)|\mu,\nu\rangle$.
The full TBA equations for the principal chiral
models are known [\Ref{ORW}]; however, for particular choices of $q$,
extending the philosophy of [\Ref{FNW}-\Ref{BNNW}],
we conjecture that only one
particle contributes to the ground-state and the infinite set of TBA
equations reduces to a single equation. The precise formulation of our
conjecture is that when
$q=\omega_a/(2\omega_a^2)$ only the unique particle with the highest
charge/mass ratio contributes to the ground-state, i.e.
the particle $|\omega_a,\omega_a\rangle$ which is highest weight state
of the multiplet $W_a$. This particle has $Q$ eigenvalue 1.
However, we exclude the
the spinor particles of SO($N$) from this conjecture.
We shall find that this proposal leads to a result which is perfectly
consistent with the perturbative calculation. Notwithstanding this, it
should be possible to prove the conjecture directly from the full TBA
equations.

The expression for the free-energy with $q=\omega_a/(2\omega_a^2)$ is then
given in terms of a quantity $\epsilon(\t)$ which satisfies the
integral equation:
$$
\epsilon(\t)-\int_{-B}^Bd\t'\phi_a(\t-\t')\epsilon(\t')=m_a\cosh\t-h.
\nfr{TBA}
The parameter $B$ is determined by the boundary condition
$\epsilon(\pm B)=0$ and the kernel is
given by
$$
\phi_a(\t)={1\over2\pi i}{d\over d\t}\ln S_{aa}(\t)=\delta(\t)-
\int_0^\infty{dx\over\pi}\cos(x\t)R_{aa}(x),
\efr
where $S_{aa}(\t)$ is the $S$-matrix element of the particle
$|\omega_a,\omega_a\rangle$ with itself \SM. Once $\epsilon(\t)$ is
known the expression for the free-energy per unit volume is
$$
\delta f(h)={m_a\over2\pi}\int_{-B}^Bd\t\,\epsilon(\t)\cosh\t.
\efr

Our problem is to solve the integral equation \TBA. In general it is
not possible to find the
solution of such an equation in closed form; however,
for comparing with the perturbative result we only need to compute the
free-energy in the ultra-violet regime $h\gg m$. In this limit a
series solution can be found using generalized Wiener-Hopf techniques
[\Ref{FNW},\Ref{HMN},\Ref{JNW}]. The first problem is to decompose the
kernel $R_{aa}(x)$:
$$
R_{aa}(x)={1\over G_+^{(a)}(x)G_-^{(a)}(x)},
\efr
where $G_\pm^{(a)}(x)$ are analytic in the upper/lower half-planes,
respectively, and $G_-^{(a)}(x)=G_+^{(a)}(-x)$. The next step in the solution
technique depends upon the form of $G_+^{(a)}(x)$. For all the principal
chiral models
$$
G_+^{(a)}(i\xi)={k'_a\over\sqrt\xi}\left\{1-b_a\xi+{\cal O}(\xi^2)\right\},
\nfr{GF}
for constants $k'_a$ and $b_a$. So in this respect
these models are of similar type to
the O($N$) sigma model rather than the fermion models. With
$G_+^{(a)}(i\xi)$ of the form \GF, [\Ref{BNNW}] gives a formula for the
first few terms in the expansion of the free-energy
$$\eqalign{
&\delta f(h)=\cr&-{h^2\over4}k^{\prime2}_a\left[\ln{h\over m_a}
+\ln\left({\sqrt{2\pi}k'_ae^{-b_a}\over
G_+^{(a)}(i)}\right)-1+
{1\over2}\ln\ln{h\over m_a}+{\cal
O}\left({1\over\ln{h\over\Lambda_{\overline{\rm MS}}}}\right)
\right].\cr}
\nfr{SMFE}

The explicit expressions for the decompositions are for ${\rm A}_r$
$$\eqalign{
G_+^{(a)}(i\xi)=&{r+1\over\sqrt{2\pi a(r+1-a)\xi}}
{\Gamma\left(1+{a\over r+1}\xi\right)
\Gamma\left(1+{r+1-a\over r+1}\xi\right)\over\Gamma(1+\xi)}\cr
&\qquad\qquad\times\exp\left\{-\xi\left({r+1-a\over r+1}\ln
{r+1-a\over r+1}+{a\over r+1}\ln{a\over r+1}\right)\right\}.\cr
}\efr
For the other algebra one finds the universal form
$$\eqalign{
G_+^{(a)}(i\xi)=&\sqrt{g\over2\pi a\xi}{\Gamma\left(1+{a\over
g}\xi\right)\Gamma\left({1\over2}+{
g-2a\over2g}\xi\right)\over\Gamma\left({1\over2}+{1\over2}\xi\right)}\cr
&\qquad\qquad\times\exp\left\{
-\xi\left({a\over g}\ln{a\over g}+{g-2a\over2g}\ln{g-2a\over2
g}-{1\over2}\ln{1\over2}\right)\right\}.\cr}
\efr
{}From these expressions we find that $k'_a$ equals $k_a$ in \KA\ and \KR\ and
for ${\rm A}_r$
$$
b_a={r+1-a\over r+1}\ln
{r+1-a\over r+1}+{a\over r+1}\ln{a\over r+1},
\efr
whilst for the other algebras
$$
b_a={a\over g}\ln{a\over g}+{g-2a\over2g}\ln{g-2a\over
2g}-{1\over2}\ln{1\over2}-{2a\over g}\ln2.
\efr

Comparing the expression \SMFE\ with the result of the perturbative
calculation \PFE\ we see that they are in perfect agreement if
$m_a\propto\sin(\pi a/g)$ which is true for all the particles
excluding the spinors of SO($N$) \MASS, and furthermore the expression for the
mass-gap has a universal form:
$$
{m\over\Lambda_{\overline{\rm MS}}}={g\over
\sqrt{\pi e}}\exp\left\{\left({2d_1+g\over
2g}\right)\ln2\right\}\sin\left({\pi\over g}\right),
\nfr{MG}
where $m$ is the mass of the vector particle, which is the lightest
particle in the theory (since without loss of generality it is only
necessary to consider ${\rm B}_r$ for $r\geq3$ and ${\rm D}_r$ for $r\geq4$).

In addition the $S$-matrix calculation implies that the universal
ratio $\beta_1/\beta_0^2=1/2$ in exact agreement with the
perturbative calculation of [\Ref{MS}] a fact first pointed out for
the SU($r+1$) theories in [\Ref{W},\Ref{PW}].
The expression for the mass-gap \MG\ reduces to that of [\Ref{BNNW}]
for ${\rm A}_r$. The explicit expressions for each group/algebra are
$$\eqalign{
{\rm SU}(r+1),{\rm A}_r:\qquad &{m\over\Lambda_{\overline{\rm MS}}}=
{r+1\over\sqrt{\pi e}}2^{3/2}\sin\left({\pi\over r+1}\right),\cr
{\rm SO}(2r+1),{\rm B}_r:\qquad &{m\over\Lambda_{\overline{\rm
MS}}}={2r-1\over\sqrt{\pi
e}}2^{(6r+1)/(4r-2)}
\sin\left({\pi\over2r-1}\right),\cr
{\rm Sp}(2r),{\rm C}_r:\qquad &{m\over\Lambda_{\overline{\rm
MS}}}={2r+2\over\sqrt{\pi e}}2^{(3r+1)/(2r+2)}
\sin\left({\pi\over 2r+2}\right),\cr
{\rm SO}(2r),{\rm D}_r:\qquad &{m\over\Lambda_{\overline{\rm
MS}}}={2r-2\over\sqrt{\pi e}}2^{(3r-1)/(2r-2)}
\sin\left({\pi\over2r-2}\right).\cr
}\efr

The fact that the perturbative result and the $S$-matrix result are
consistent provides strong grounds for believing that our conjecture about
the structure of the ground-states is correct. As has been pointed out
in [\Ref{BNNW}], the fact that the TBA calculation reproduces the
universal part of the beta-function $\beta_1/\beta_0^2$ is a highly
non-trivial fact. In addition if the $S$-matrix were modified with CDD
factors then the thermodynamics would be drastically altered and the
perfect agreement with the perturbative result would be destroyed.
It would be interesting
to compare these results with lattice simulations.

I would like to thank Tim Morris and Michel Bauer for very useful
discussions.

\references

\beginref
\Rref{ORW}{E. Ogievetsky, N. Reshetikhin and P. Wiegmann, Nucl. Phys.
{\bf B280} (1987) 45}
\Rref{FNW}{P. Forg\'acs, F. Niedermayer and P. Weisz, Nucl. Phys. {\bf
B367} (1991) 123}
\Rref{MS}{A. McKane and M. Stone, Nucl. Phys. {\bf B163} (1980) 169}
\Rref{Y}{S.-K. Yang, Nucl. Phys. {\bf B267} (1986) 290}
\Rref{HN}{P. Hasenfratz and F. Niedermayer, Phys. Lett. {\bf B245}
(1990) 529}
\Rref{BNNW}{J. Balog, S. Naik, F. Niedermayer and P. Weisz, Amsterdam
Lattice (1992) 232}
\Rref{HMN}{P. Hasenfratz, M. Maggiore and F. Niedermayer, Phys. Lett.
{\bf B245} (1990) 522}
\Rref{JNW}{G. Japaridze, A. Nersesyan and P. Wiegmann, Nucl. Phys.
{\bf B230} (1984) 511}
\Rref{W}{P.B. Wiegmann, Phys. Lett. {\bf B141} (1984) 217}
\Rref{PW}{A. Polyakov and P.B. Wiegmann, Phys. Lett. {\bf B131} (1983)
121}
\Rref{TH}{T.J. Hollowood, ``{\sl The analytic structure of
trigonometric $S$-matrices\/}'', CERN preprint CERN-TH.6888/93, {\tt
hep-th/9305042}, {\it to appear in\/}: Nucl. Phys. {\bf B}}
\endref
\ciao